\documentclass[12pt]{iopart}
\usepackage{iopams,epsfig}
\newcommand{\myext}{ps}
\newcommand{\mypicwidth}{11cm}
\newcommand{\mypicwidthsmall}{9cm}

\newcommand{\F}{{\mathcal F}}

\newcommand{\exc}{{\rm exc}}
\newcommand{\rv}{{\bf r}}
\newcommand{\xv}{{\bf x}}

\newcommand{\Gozdz}{G\'o\'zd\'z}

\begin{document}

\letter{Rosenfeld functional for non-additive hard spheres}

\author{Matthias Schmidt\footnote{On leave from:  
    Institut f\"ur Theoretische Physik II,
    Heinrich-Heine-Universit\"at D\"usseldorf, Universit\"atsstra\ss e 1,
    D-40225 D\"usseldorf, Germany.}}
\address{
    Soft Condensed Matter Group, 
    Debye Institute, Utrecht University, Princetonplein 5,
    3584 CC Utrecht, The Netherlands.
}

\begin{abstract}
  The fundamental measure density functional theory for hard spheres
  is generalized to binary mixtures of arbitrary positive and moderate
  negative non-additivity between unlike components. In bulk the
  theory predicts fluid-fluid phase separation into phases with
  different chemical compositions. The location of the accompanying
  critical point agrees well with previous results from simulations
  over a broad range of non-additivities and both for symmetric and
  highly asymmetric size ratios. Results for partial pair correlation
  functions show good agreement with simulation data.
\end{abstract}
\submitto{\JPCM on 18 June 2004, revised version 30 June 2004}
\vspace{5mm}

Density-functional theory (DFT) is a powerful approach to study
equilibrium properties of inhomogeneous systems, including dense
liquids and solids of single- and multi-component substances
\cite{evans92}. Its practical applicability depends on the quality of
the approximation to the central object, the (Helmholtz) excess free
energy functional arising from the interparticle interactions. The
specific model of additive hard sphere mixtures constitutes the
reference system par excellence to describe mixtures governed by
steric repulsion, and Rosenfeld's fundamental-measure theory (FMT)
\cite{rosenfeld89,RSLTlong,tarazona00,cuesta02} is arguably the best
available approximation to tackle inhomogeneous situations. A rapidly
increasing number of applications to interesting physical problems can
be witnessed \cite{jpcm2002Rosenfeld}.

The more general non-additive hard sphere mixture is defined through
pair potentials between particles of species $i$ and $j$, given as
$V_{ij}(r)=\infty$ for $r<\sigma_{ij}$ and 0 otherwise, where $r$ is
the center-center distance between the two particles, and
$\sigma_{ij}$ is the distance of minimal approach between particles of
species $i$ and $j$. In a binary mixture non-additivity is measured
conventionally through the parameter
$\Delta=2\sigma_{12}/(\sigma_{11}+\sigma_{22})-1$. The physics of
non-additive hard sphere mixtures is considerably richer than that of
the additive case. In particular the case of $\Delta>0$ is striking,
as small values of $\Delta$ are known to be already sufficient to
induce stable fluid-fluid demixing into phases with different chemical
compositions (for recent studies see e.g.\ references
\cite{dijkstra98II,louis00III,gozdz03,jagannathan03}).

The treatment of general non-additivity is elusive in the FMT
framework. The author is aware of successful studies only in four
special cases: First, for the Asakura-Oosawa-Vrij (AOV) model
\cite{asakura54,vrij76}, where species 1 represents colloidal hard
spheres and species 2 (with $\sigma_{22}=0$) represents
non-interacting polymer coils with radius of gyration equal to
$\sigma_{12}-(\sigma_{11}/2)$, an excess free energy functional was
given \cite{schmidt00cip}. Second, a free energy functional for the
Widom-Rowlinson (WR) model, where $\sigma_{11}=\sigma_{22}=0$, was obtained
\cite{schmidt01wr}. Third, the depletion potential between two big
spheres immersed in a sea of smaller spheres was obtained through
``Roth's trick'' of working on the level of the one-body direct
correlation functional
\cite{roth01nonadditive,roth01nonadditiveII,louis01depl}. In this case
the functional for the additive case is sufficient to obtain results,
but the approach is limited to small concentration of big
spheres. Fourth, in Lafuente and Cuesta's FMT for lattice hard core
models, due the an odd-even effect of the particle sizes (measured in
units of lattice constants), non-additivity of the size of one lattice
spacing arises \cite{lafuente02}. This effect, however, is specific to
lattice models and vanishes in the continuum limit.

The aim of the present letter is to generalize FMT for hard spheres to
the case of general positive and moderate negative non-additivity and
arbitrary size asymmetry. The proposed extended framework
accommodates, in the respective limits, the Rosenfeld functional for
additive hard sphere mixtures \cite{rosenfeld89}, the DFT for the
extreme non-additive AOV case \cite{schmidt00cip}, and the exact
virial expansion up to second order in densities. The structure of the
theory, however, goes qualitatively beyond that of either limit.

\begin{figure}
  \begin{center}
    \includegraphics[width=\mypicwidthsmall]{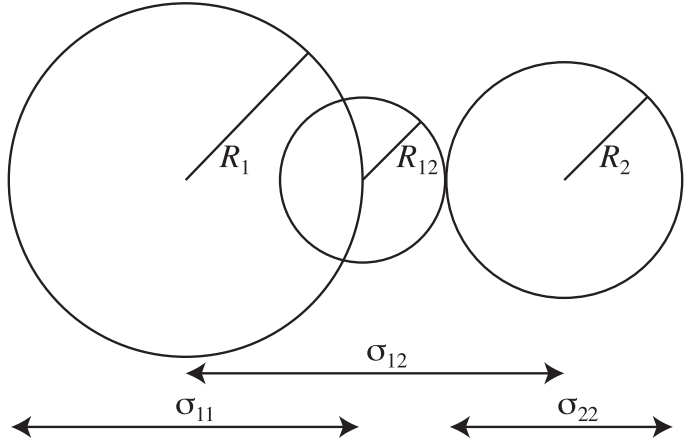}
    \caption{Illustration of the relevant length scales. The hard core
      interaction distances $\sigma_{ij}$ between pairs of particles
      of species $ij=11,12$ and 22 are related to radii through
      $\sigma_{11}=2R_1$, $\sigma_{12}=R_1+R_{12}+R_2$ and
      $\sigma_{22}=2R_2$, respectively. The spheres of radii $R_1$ and
      $R_2$ represent the weight functions $w_\alpha^{(1)}$ and
      $w_\beta^{(2)}$, respectively, and can be viewed as ``true''
      particle shapes.  The sphere of radius $R_{12}$ represents the
      kernel $K_{\alpha\beta}^{(12)}$ being a mere construct to
      generate the correct hard core distance $\sigma_{12}$ between
      species 1 and~2.}
    \label{FIGillustration}
\end{center}
\end{figure}

The excess (over ideal) Helmholtz free energy functional is expressed as
\begin{equation}
\fl\quad  \F_\exc[\rho_1,\rho_2] = k_BT\int d\xv d\xv' 
  \sum_{\alpha,\beta=0}^3 K^{(12)}_{\alpha\beta}(|\xv-\xv'|)
  \Phi_{\alpha\beta}\left(\{n_\nu^{(1)}(\xv)\},\{n_\tau^{(2)}(\xv')\}\right),
  \label{EQfexc}
\end{equation}
where $\rho_i(\rv)$ is the one-body density distributions of species
$i=1,2$ dependent on position~$\rv$, $k_BT$ is the thermal energy,
$\Phi_{\alpha\beta}$ for $\alpha,\beta=0,1,2,3$ is the free energy
density depending on the sets of weighted densities $\{n_\nu^{(i)}\}$
for $i=1,2$, and the kernels $K_{\alpha\beta}^{(12)}(r)$ are a means
to control the range of non-locality between unlike components and
depend solely on distance $r$.  The weighted densities are built in
the usual way~\cite{rosenfeld89} through convolution of the respective
bare density profile with appropriate weight functions:
\begin{equation}
  n_\nu^{(i)}(\xv) = \int d\rv \rho_i(\rv) w_\nu(|\xv-\rv|,R_i), 
  \quad i=1,2,
  \label{EQweightedDensities}
\end{equation}
where $\nu=0,1,2,3$ labels the type of weight function, and
$R_i=\sigma_{ii}/2$ is the particle radius of species $i=1,2$.  The
(fully scalar) Kierlik-Rosinberg form \cite{kierlik90,phan93} of the
$w_\nu(r,R)$ is used in the following, as this renders the
determination of the $K_{\alpha\beta}^{(12)}(r)$ more straightforward.
The $w_\nu(r,R)$ are
\begin{equation}
\eqalign{
  w_0 &= -\delta''(R-r)/(8\pi)+\delta'(R-r)/(2\pi r),\label{EQwFirst}\\
  w_1 &= \delta'(R-r)/(8\pi),\\
  w_2 &= \delta(R-r),\\
  w_3 &= \Theta(R-r),\label{EQwLast}}
\end{equation}
where $R=R_i$, the prime denotes the derivative w.r.t.\ the argument,
$\delta(\cdot)$ is the Dirac distribution, and $\Theta(\cdot)$ is the
step function. Alternatively, in Fourier space the weight functions
are $\tilde w_\alpha(k,R) = 4\pi\int dr w_\alpha(r,R)\sin(kr)r/k$ and
given as
\begin{eqnarray}\eqalign{
  \tilde w_0 &= c+(kRs/2),\\
  \tilde w_1 &= (kRc+s)/(2k),\\
  \tilde w_2 &= 4\pi Rs/k,\\
  \tilde w_3 &= 4\pi(s- kRc)/k^3,}
\label{EQweightsInFourierSpace}
\end{eqnarray}
with the abbreviations $s=\sin(kR)$ and $c=\cos(kR)$. The kernels
$K_{\alpha\beta}^{(12)}(r)$ in (\ref{EQfexc}) can be viewed as
$\alpha\beta$-components of a second-rank tensor
\begin{equation}
\hat{\bf K}^{(12)}(r)= \left(
  \begin{tabular}{cccc}
      $w_{3}$ & $w_{2}$ & $w_{1}$ & $w_0$\\
      $w_{2}$ & $w_{1}^\dag$ & $w_0^\dag$ & $w_{-1}$ \\
      $w_{1}$ & $w_0^\dag$ & $w_{-1}^\dag$ & $w_{-2}$ \\
      $w_0$ & $w_{-1}$ & $w_{-2}$ & $w_{-3}$
  \end{tabular}
  \right),\\
\end{equation}
where indexing is such that the top row contains
$K_{00}^{(12)},\ldots, K_{03}^{(12)}$, etc, and $\dag$ distinguishes
different elements.  All $K_{\alpha\beta}^{(12)}(r)$ possess a range
of $R_{12}=\sigma_{12}-(\sigma_{11}+\sigma_{22})/2$, i.e.\ vanish for
values of $r$ beyond that distance (see figure \ref{FIGillustration}
for an illustration of the length scales). The dimension of
$K_{\alpha\beta}^{(12)}$ is $(\rm length)^{-\alpha-\beta}$, and hence
the dimension of $w_\gamma$ is $(\rm length)^{\gamma-3}$. The elements
of $\hat{\bf K}^{(12)}$ are defined, with $R=R_{12}>0$, through
(\ref{EQwFirst}), and furthermore
\begin{eqnarray}\eqalign{
   w_1^\dag &= \delta'(R-r),\\
   w_0^\dag &= \delta''(R-r)/(8\pi),\\
   w_{-1}^\dag &= \delta^{(3)}(R-r)/(64\pi^2),\\
   w_{-1} &= \delta^{''}(R-r)/(2\pi r)-\delta^{(3)}(R-r)/(8\pi),\\
   w_{-2} &= \delta^{(3)}(R-r)/(16\pi^2r)-\delta^{(4)}(R-r)/(64\pi^2),\\
   w_{-3} &= -\delta^{(4)}(R-r)/(8\pi^2r)+\delta^{(5)}(R-r)/(64\pi^2),}
\end{eqnarray}
with the derivatives $\delta^{(\gamma)}(x)=d^\gamma \delta(x)/d
x^\gamma$ for $\gamma=3,4,5$. Again we also give the Fourier space
representation [being together with (\ref{EQweightsInFourierSpace})
also valid for $R=R_{12}<0$], which reads
\begin{eqnarray}\eqalign{
  \tilde w_1^\dag &= 4\pi(kRc+s)/k,\\
  \tilde w_0^\dag &= c-(kRs/2),\\
  \tilde w_{-1}^\dag &= -(k^2Rc+3ks)/(16\pi),\\
  \tilde w_{-1} &= (k^2Rc-ks)/2,\\
  \tilde w_{-2} &= -k^3Rs/(16\pi),\\
  \tilde w_{-3} &= (k^4Rc - 3k^3s)/(16\pi).}
\label{EQkernelsInFourierSpace}
\end{eqnarray}
In order to express the dependence of the free energy density,
$\Phi_{\alpha\beta}$ in equation (\ref{EQfexc}), on the weighted
densities (\ref{EQweightedDensities}) we introduce ansatz functions
$A_{\alpha\gamma}^{(i)}$ for species $i=1,2$ that possess the
dimension of $(\rm length)^{\alpha-3}$ and the order $\gamma$ in
density (i.e.\ contain $\gamma$ factors $n_\tau^{(i)}$). Explicit
expressions for the non-vanishing terms are
\begin{eqnarray}
  A_{01}^{(i)} = n_0^{(i)}, \label{EQcoefficientA01} \quad
  A_{02}^{(i)} = n_1^{(i)}n_2^{(i)},\quad
  A_{03}^{(i)} = \left(n_2^{(i)}\right)^3/(24\pi),\\
  A_{11}^{(i)} = n_1^{(i)},\quad
  A_{12}^{(i)} = \left(n_2^{(i)}\right)^2/(8\pi),\quad
  A_{21}^{(i)} = n_2^{(i)},\quad
  A_{30}^{(i)} = 1.
\end{eqnarray}
The excess free energy density is then constructed as
\begin{equation}
  \Phi_{\alpha\beta} = \sum_{\gamma=0}^6\sum_{\gamma'=0}^3
  A_{\alpha\gamma'}^{(1)} A_{\beta(\gamma-\gamma')}^{(2)}
  \varphi_{\rm 0d}^{(\gamma)}\left(n_3^{(1)}+n_3^{(2)}\right),
  \label{EQphiOfWeightedDensities}
\end{equation}
where $\varphi_{0d}^{(\gamma)}(\eta)\equiv d^\gamma
\varphi_{0d}(\eta)/d\eta^\gamma$ is the $\gamma$th derivative of the
zero-dimensional excess free energy as a function of the average
occupation number $\eta$ \cite{RSLTlong},
$\varphi_{0d}(\eta)=(1-\eta)\ln(1-\eta)+\eta$, and
$\varphi_{0d}^{(0)}(\eta)\equiv\varphi_{0d}(\eta)$ for $\gamma=0$.
The specific form (\ref{EQphiOfWeightedDensities}) ensures both that
the terms in the sum in (\ref{EQfexc}) possesses the correct dimension
of $(\rm length)^{-6}$ and that the prefactor of $\varphi_{\rm
0d}^{(\gamma)}$ in (\ref{EQphiOfWeightedDensities}) is of the total
order $\gamma$ in densities, as is common in FMT. This completes the
prescription for the functional; a full account of all details, also
for multi-component mixtures and for lower spatial dimensionality,
will be given elsewhere.

Here we discuss some of the properties of the theory. For small
densities it is straightforward to show that the correct virial
expansion up to second order in densities is obtained, $F_{\rm exc}\to
-\sum_{ij}\int d^3rd^3r'f_{ij}(|\rv-\rv'|)\rho_i(\rv)\rho_j(\rv')/2$,
where the Mayer functions, $f_{ij}(r)=\exp(-V_{ij}(r)/(k_BT))-1$, are
recovered through
\begin{eqnarray}
  f_{12} = -\sum_{\alpha\beta=0}^3
   w_\alpha^{(1)}\ast K_{\alpha\beta}^{(12)}\ast w_\beta^{(2)}, \quad
  f_{ii} = -\sum_{\alpha=0}^3 w_\alpha^{(i)} \ast w_{3-\alpha}^{(i)},\; i=1,2,
\end{eqnarray}
where $\ast$ denotes the convolution, $g(\rv)\ast h(\rv)=\int d^3r'
g(\rv')h(\rv-\rv')$. In the limit of an additive mixture, $\Delta\to
0$ and hence $R_{12}\to 0$, one finds that $K_{\alpha\beta}(x)\to 0$
if $\beta\neq 3-\alpha$ and $K_{\alpha(\alpha-3)}(x)\to\delta(x)$
otherwise. This leads to a cancellation of one spatial integration in
(\ref{EQfexc}) and yields the Rosenfeld functional for a binary
additive hard sphere mixture \cite{rosenfeld89} with radii $R_1$ and
$R_2$. In the AOV limit, $R_2\to 0$, one finds that $n_\alpha^{(2)}\to
0$ if $\alpha\neq 0$, and $n_0^{(2)}\to\rho_2$ otherwise. The
integration over $\xv'$ in (\ref{EQfexc}) together with the kernel
$K_{\alpha\beta}(|\xv-\xv'|)$ and the fact that the density
$n_{0}^{(2)}(\xv')=\rho_2(\xv')$ appears linearly in
$\Phi_{\alpha\beta}$, see $A_{01}^{(2)}$ in (\ref{EQcoefficientA01}),
plays the same role that building weighted densities for the polymer
species in the AOV case does. The resulting functional is equal to
that for the AOV model \cite{schmidt00cip}.  However, in the WR limit,
in contrast to \cite{schmidt01wr}, terms higher than on the second
virial level vanish. For $\Delta=-1$ the two species decouple, and
${\cal F}_{\rm exc}[\rho_1,\rho_2]={\cal F}_{\rm exc}[\rho_1]+ {\cal
F}_{\rm exc}[\rho_2]$ which is {\em not} obeyed by the present
approximation, limiting its applicability to small values of
$\Delta<0$.

\begin{figure}
  \begin{center}
    \includegraphics[width=\mypicwidth]{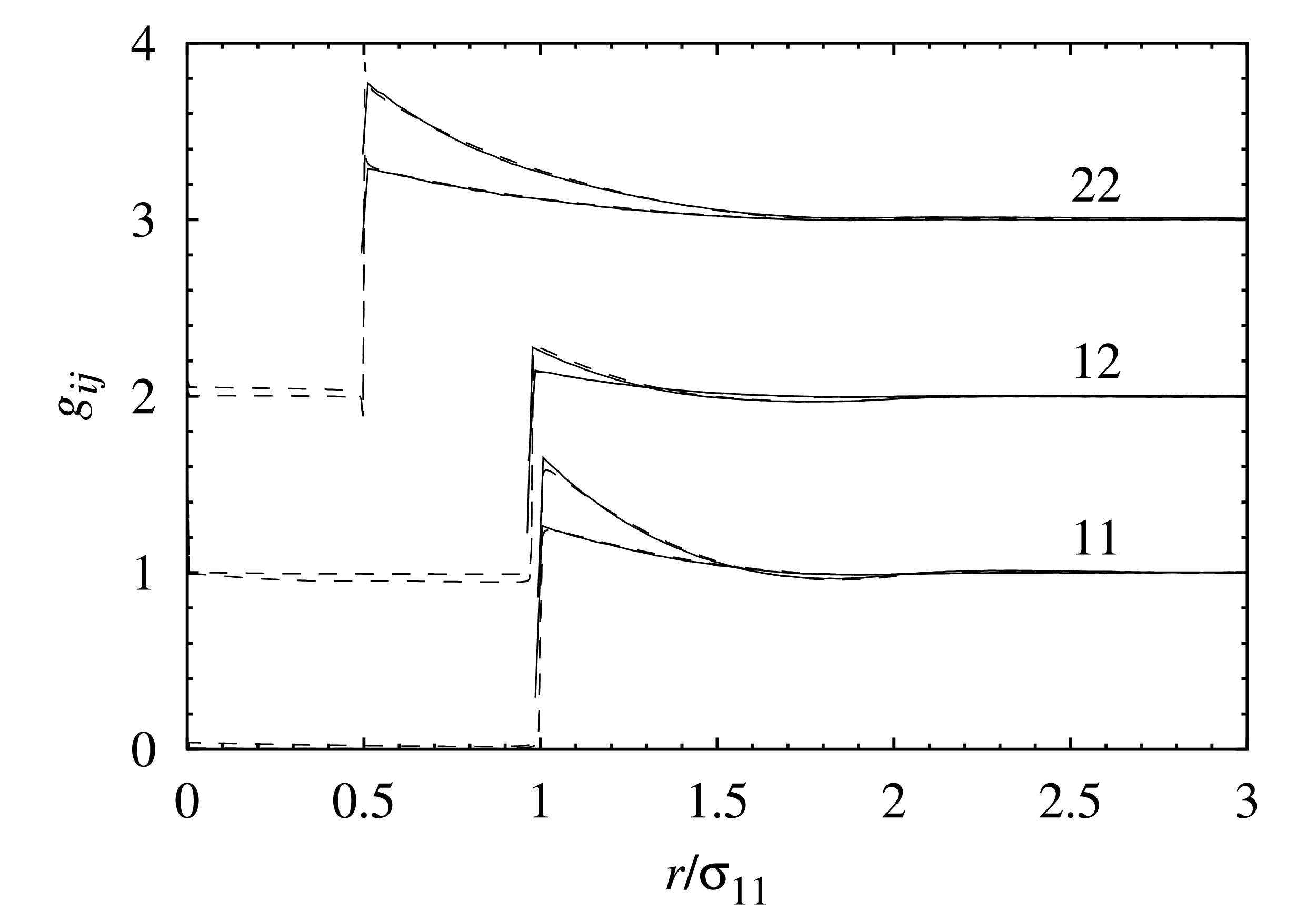}
    \caption{Partial pair correlation functions, $g_{ij}(r)$, between
      species $ij=11,12$ and 22 (as indicated), as a function of the
      scaled distance $r/\sigma_{11}$, as obtained from the present
      DFT using the OZ route (dashed lines) and from MC simulation
      (solid lines). Results for $g_{12}$ ($g_{22}$) are shifted
      upwards by one (two) units for clarity. Parameters are
      $\sigma_{22}/\sigma_{11}=0.5$, $\Delta=0.3$, $\eta_2=\eta_1/8$
      and $\eta_1=0.05$ (lower), 0.1 (upper). For comparison, the
      theoretical critical point is located at $\eta_1=0.118,
      \eta_2=0.0321$.}
    \label{FIGgofr}
\end{center}
\end{figure}

We next turn to an investigation of bulk properties of the theory.  To
assess structure, direct correlation functions can be obtained via
\begin{equation}
 c_{ij}^{(2)}(|\rv-\rv'|) = \left.-\frac{\delta^2 \beta \F_\exc}
 {\delta \rho_i(\rv) \delta \rho_j(\rv')}\right|_{\rho_1,\rho_2=\rm const},
\end{equation}
which can be shown to feature Percus-Yevick (PY) like behavior:
$c_{ij}^{(2)}(r>\sigma_{12})=0$.  Inverting the Ornstein-Zernike (OZ)
relations permits to calculate partial structure factors, $S_{ij}(k)$,
and partial pair correlation functions, $g_{ij}(r)$. We have carried
out Monte Carlo (MC) computer simulations in the canonical ensemble
with 1024 particles and $10^5$ MC moves per particle; histograms of
all distances between particles yield benchmark results for
$g_{ij}(r)$. We have chosen an intermediate size ratio of
$\sigma_{22}/\sigma_{11}=0.5$ and have considered various values of
$\Delta$ from $-0.3$ to 0.5 and a range of statepoints characterized
by packing fractions, $\eta_i=\pi\rho_i \sigma_{ii}^3/6$ for
$i=1,2$. For $\Delta=0$, the current DFT reproduces the solution of
the PY integral equation, as the functional reduces to the Rosenfeld
case, which is known to yield the same $c_{ij}^{(2)}(r)$ as the PY
approximation.  Results for the representative case $\Delta=0.2$ at
two different statepoints are shown in figure \ref{FIGgofr}. The core
condition, $g_{ij}(r<\sigma_{ij})=0$, is only approximately fulfilled,
but the overall agreement between results from theory and simulation
is quite remarkable.

In principle one could envisage that this approach permits to study
the depletion potential, $V_{\rm depl}^{(11)}(r)$, between particles
of species 1 being generated by the immersion into a ``sea'' of
particles 2 through $V_{\rm depl}^{(11)}(r)=-k_BT\ln g_{11}(r)$ for
$\rho_1\to 0$, and $\rho_2=\rm const$. However, for the (relevant)
case of small size ratios (e.g.\ $\sigma_{22}/\sigma_{11}\sim 0.1$,
see \cite{roth01nonadditive,roth01nonadditiveII}) already in both
limits of additive hard spheres and the AOV model the results are only
of rather moderate accuracy, underestimating the strength of the
depletion attraction \cite{schmidt00cip}, similar to results from the
PY approximation. However, results from the present theory obtained
through the OZ route (not shown) cross over smoothly between the
additive hard sphere case and the AOV case, similar to the correct
behavior \cite{roth01nonadditive,roth01nonadditiveII}.  Hence one can
conclude that the pair structure predicted by the current DFT is
similar to that of the PY approximation. This is a remarkable
property, and one can anticipate test-particle calculations to yield
superior results.

Evaluating (\ref{EQfexc}) at constant density fields yields an
analytic expression for the bulk excess free energy for fluid states,
$F_{\rm exc}={\cal F}_{\rm exc}[\rho_1{\rm =const},\rho_2{\rm
=const}]$.  The total Helmholtz free energy is then $F=F_{\rm
exc}+k_BTV\sum_{i=1,2} \rho_i[\ln(\rho_i\Lambda_i^3)-1]$, where
$\Lambda_i$ is the (irrelevant) de Broglie wavelength of species $i$,
and $V$ is the system volume. Via Taylor expanding $F_{\rm exc}$ in
both densities one can show that it features the exact second virial
coefficients (consistent with the correct incorporation of $f_{ij}(r)$
on the second virial level) and also the exact third virial
coefficients (see e.g.\ \cite{dijkstra98II})
provided $2\sigma_{12}>\max(\sigma_{11},\sigma_{22})$.

\begin{figure}
  \begin{center}
    \includegraphics[width=\mypicwidth]{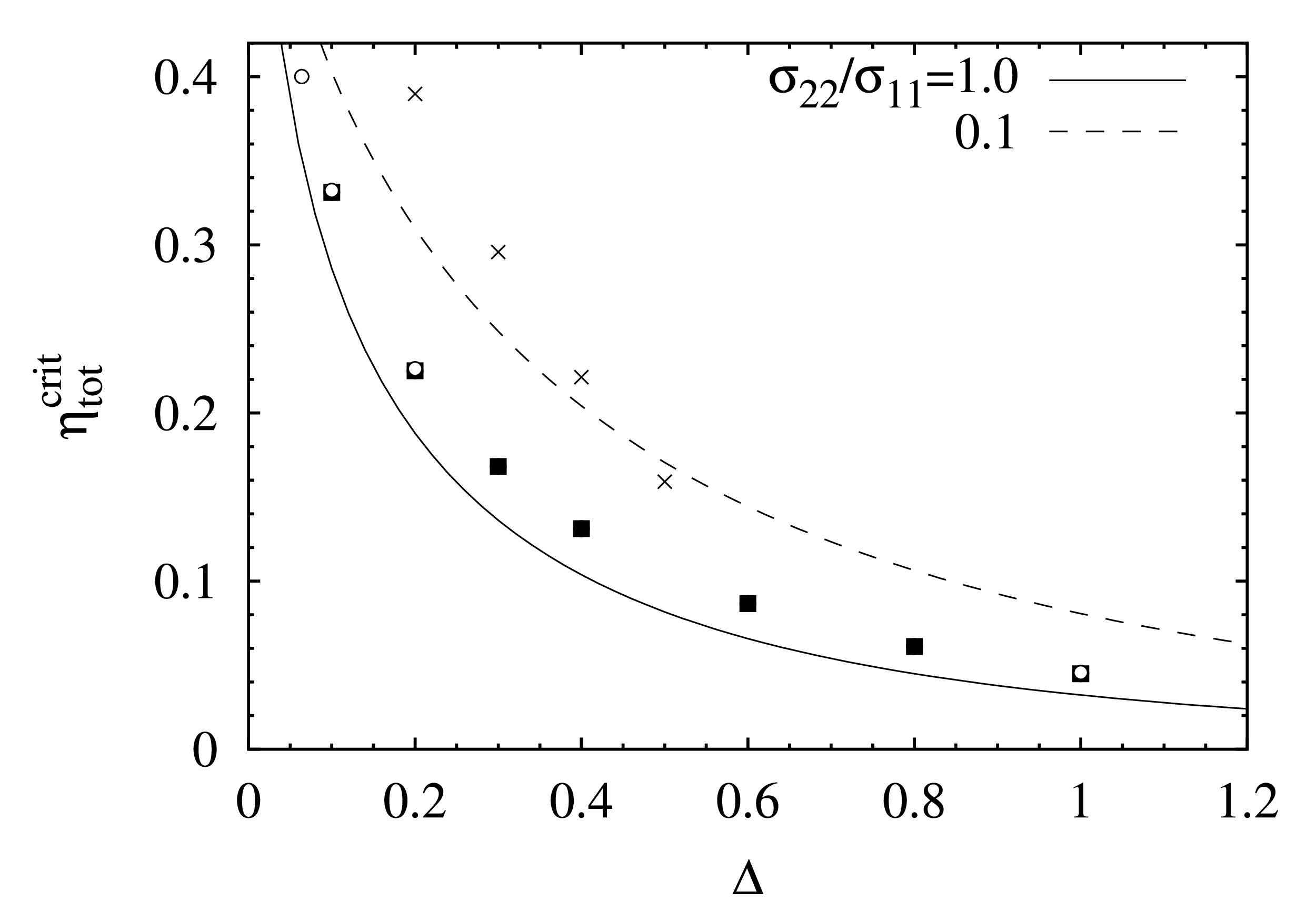}
    \caption{The total packing fraction at the critical point,
      $\eta_{\rm tot}^{\rm crit}$, where $\eta_{\rm
      tot}=\eta_1+\eta_2$, for a non-additive binary hard sphere
      mixture as a function of the non-additivity parameter
      $\Delta$. Shown are results from the present DFT (lines) and
      from simulations (symbols) for the symmetric case,
      $\sigma_{22}/\sigma_{11}=1$, by \Gozdz~\cite{gozdz03} (filled
      squares) and by Jagannathan and Yethiraj \cite{jagannathan03}
      (open circles), as well as for the highly asymmetric case of
      $\sigma_{22}/\sigma_{11}=0.1$ by Dijkstra \cite{dijkstra98II}
      (crosses).}
    \label{FIGcrit}
\end{center}
\end{figure}

\begin{figure}
  \begin{center}
    \includegraphics[width=\mypicwidth]{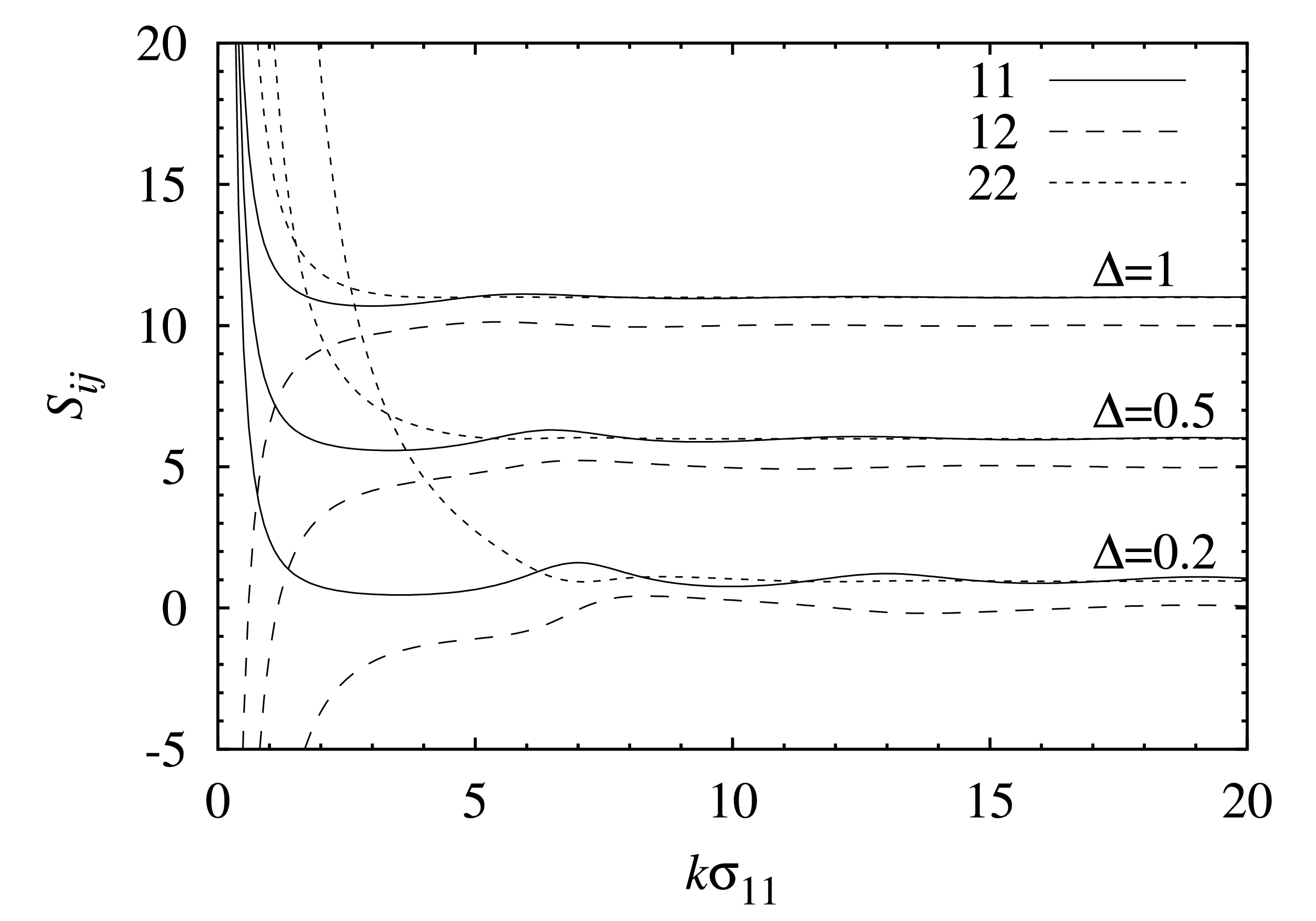}
    \caption{Partial structure factors, $S_{ij}(k)$ for $ij=11,12,22$
      (as indicated), as a function of $k\sigma_{11}$ at the
      fluid-fluid critical point for size ratio
      $\sigma_{22}/\sigma_{11}=0.1$ and non-additivity $\Delta=0.2,
      0.5, 1$. The results for $\Delta=0.5$ (1) are shifted upwards by
      5 (10) units for clarity.}
    \label{FIGsij}
\end{center}
\end{figure}

The fluid-fluid demixing spinodal can be obtained from (numerical)
solution of $|\partial^2 (F/V)/\partial\rho_i\partial\rho_j|=0$, and
the location of the critical point can be determined from minimizing
one of the chemical potentials, $\mu_1$ or $\mu_2$, along the
spinodal. Such results are compared in figure \ref{FIGcrit} to those
from simulations for $\sigma_{11}=\sigma_{22}$, performed in the
semi-grand ensemble by Jagannathan and Yethiraj \cite{jagannathan03}
and by \Gozdz~\cite{gozdz03}, the latter study including a finite size
analysis, for a variety of non-additivities ranging from
$\Delta=0.1-1$.  For the highly asymmetric case of
$\sigma_{22}=\sigma_{11}$ results from Gibbs ensemble simulations were
obtained by Dijkstra \cite{dijkstra98II}.  For both size ratios the
strong decrease of the total critical packing fraction with increasing
values of $\Delta$, as well as the overall functional dependence are
very well described by the theory. However, the precise value at given
$\Delta$ is underestimated. This behavior is not uncommon for
mean-field like theories and is also present in the AOV case.
A benefit of working on the level of the density functional is that
the structure is consistent with the free energy. In figure
\ref{FIGsij} partial structure factors are shown for a range of values
of $\Delta$ evaluated at the fluid-fluid critical point obtained from
the free energy, and indeed $S_{ij}(k\to 0)\to\pm \infty$.

In conclusion, having demonstrated the good accuracy of the
predictions of the current theory for bulk fluid properties of the
non-additive hard sphere mixture, we are confident that it is well
suited to study interesting and relevant interfacial situations, like
the structure and tension of interfaces between demixed phases,
wetting at substrates \cite{brader02swet} and more.  Note that any
colloidal mixtures interacting with soft repulsive forces, as e.g.\
present in charge-stabilized dispersions, can be mapped (e.g.\ by the
Barker-Henderson procedure) onto an effective non-additive hard sphere
system. Hence one can anticipate experimental consequences of the
structure and phase separation predicted by the present theory.  The
treatment of freezing \cite{louis00III} requires additional
contributions to the free energy functional
\cite{RSLTlong,tarazona00}.
\\
\noindent
H.\ L\"owen, R.\ Evans, R.\ Blaak and K. Jagannathan are thanked for
useful comments.  Support by the SFB TR6 of the DFG is
acknowledged. This work is part of the research program of FOM, that
is financially supported by the NWO.

\end{document}